\documentclass[aps,prl,amsfonts,twocolumn,showpacs,floatfix]{revtex4}

\usepackage{graphicx}
\usepackage{amsmath}
\usepackage{bm}

\begin{document}
\title{Variable range  cotunneling and  conductivity of a granular metal}

\author{M. V. Feigelman}
\affiliation{L. D. Landau Institute
for Theoretical Physics, Moscow 119334, Russia}

\author{A. S. Ioselevich}
\affiliation{L. D. Landau Institute for Theoretical Physics,
Moscow 119334, Russia}

\date{\today}

\begin{abstract}
The Efros-Shklovskii  law  for the conductivity of granular metals
is interpreted as a result of a  variable range cotunneling
process. The cotunneling between distant resonant grains is
predominantly elastic at low $T \leq T_c$, while it is inelastic
(i.e., accompanied by creation of electron-hole pairs on a string
of intermediate non-resonant grains) at $T \geq T_c$. The
corresponding E-S temperature $T_{ES}$ in the latter case is
slightly (logarithmically) $T$-dependent. The magnetoresistance in
the two cases is different: it may be relatively strong and
negative at $T \ll T_c$, while  at $T>T_c$ it is suppressed due to
inelastic processes  which destroy  the interference.

\end{abstract}

\maketitle

{\it Introduction.}
 The low-temperature conductivity of most granular metals (both
 three-dimensional samples and thin films) exhibits a typical insulating behavior,
 characterized by the  Efros-Shklovskii law
\begin{eqnarray}
\sigma\sim e^{-\sqrt{T_{ES}/T}}, \label{esl}
\end{eqnarray}
In the samples with low room-temperature conductivity this law is
observed in the whole range of $T$ (from room temperature down to
liquid helium temperatures)~\cite{exp1,exp2,exp3}. Two ingredients
are known to be necessary ~\cite{EfrosShklovskii75,ES} for the
existence of the behavior \eqref{esl} in usual doped
semiconductors with localized impurity centers: (i) soft "Coulomb
gap" in the electron density of states, and (ii) long-range
electron tunnelling between distant centers of their localization.
The original idea on the building of the Coulomb gap as presented
in Ref.\cite{EfrosShklovskii75} was recently adapted to granular
arrays in Ref.~\cite{ZhangShklovskii}. It was argued that the
principal source for this gap is random background charges and the
physical mechanisms behind these charges were discussed in detail.
The feature (ii) is quite natural for doped semiconductors as it
follows from exponential decay of wave-functions of localized
electrons. It is much less trivial for granular media where each
grain is typically connected by tunnel junctions to its nearest
neighbors only, therefore the very origin of long-range tunnelling
needs some special explanation.

In this Letter we demonstrate that the  variable range hopping in
a granular metal  involves the so called co-tunneling
process~\cite{AverinNazarov90} (either elastic or inelastic). The
elastic cotunneling is effective, if the temperature is low enough
(namely, for $T<T_c\sim {\cal L}^{-1}(\delta \cdot E_C)^{1/2}$,
where $\delta$ is the characteristic level spacing in the grains,
$E_C$ is the characteristic charging energy, and ${\cal L}\sim 10$
is a large logarithmic factor, see below). At $T>T_c$ the
conductivity is dominated by inelastic cotunneling processes. We
directly show this for the case of the granular metal with poor
room-temperature conductivity (small intergrain conductances
$g\equiv  (h/e^2R)\ll 1$). We expect that the same is also true
for samples with moderately good conductivity (cf.
e.g.~\cite{ET02,FIS04}) with properly renormalized $T_{ES}$, but
this more delicate issue will be discussed in a separate
publication.  Experimentally, relative role of elastic vs
inelastic cotunnelling processes could be detected by the presence
of noticeable low-field magnetoresistance in the hopping regime:
while elastic cotunnelling is expected to lead to negative
magnetoresistance like it was predicted for doped
semiconductors~\cite{magnet1,magnet2,magnet3,magnet4}, inelastic
cotunnelling is intrinsically incoherent and the whole effect of
magnetic field upon conductivity is localized within individual
grains and can, therefore, be only  observed in very high fields
$H\gtrsim 10 T$ .

Co-tunnelling as a key mechanism of low-temperature charge
transfer was proposed~\cite{AverinNazarov90} and extensively
discussed (cf. e.g. ~\cite{GlazmanReview} for the review) with
regard to transport via quantum dots.  Quantum dot situated
between two bulk metal reservoirs is characterized by its charing
energy $E_C$ and dimensionless conductances $g_{R,L}$.
Semiclassical "orthodox theory" of Coulomb
blockade~\cite{orthodox} predicts exponential suppression of
conductance through quantum dot at temperatures $T \ll E_C$, i.e.
$G_{\rm orth} \propto \min(g_L,g_R)\exp(-E_C/T)$, due to low
probability of creation of real state of the dot with extra
electron charge.  Co-tunnelling process, on the other hand, occurs
in the next order of perturbation theory in (small) tunnelling
amplitudes $t_{R,L}$, but does not contain exponential suppression
factor $\exp(-E_C/T)$ since the dot state with extra charge occurs
as {\it virtual state} only. One should distinguish two kinds of
co-tunnelling processes, elastic and inelastic ones.  Elastic
process occurs when tunnelling {\it in} and {\it out} of the dot
deals with the same intra-dot electron eigenstate $\alpha$, thus
it leaves the dot in exactly the same quantum state as it was
before it. On the contrary, inelastic co-tunnelling leaves behind
it an excited electron-hole pair (since one electron tunnels {\it
in} the dot and populates some eigenstate $h$, whereas another
electron tunnels {\it out} of the dot from another, $p$-th
eigenstate). Elastic co-tunnelling contribution to the conductance
 scales as $G_{\rm el} \propto g_Lg_R \delta/E_C$,
 whereas inelastic co-tunnelling contribution is
 $G_{\rm inel} \propto g_Lg_R  (T/E_C)^2$, cf.~\cite{AverinNazarov90,GlazmanReview}.
 Thus, upon temperature decrease first the inelastic co-tunnelling prevails over
 classical conduction given by $G_{\rm orth}$,
 and then at $T_* \sim \sqrt{E_C\delta}$ it gives way to the
 elastic co-tunnelling. Below we generalize the above ideas to the situation
 of variable-range hopping in granular arrays.

In the spirit of the standard variable range hopping (VRH) theory,
we consider a charge transfer between two distant grains $i$ and
$j$ with anomalously small energies $\varepsilon_i \,
,\varepsilon_j\ll E_C$ of the "charged ground state" (i.e. the
states with an extra electron, or an extra hole; for explicit
definition of the energy $\varepsilon$ and detailed discussion of
the corresponding "density of ground states" see
\cite{ZhangShklovskii}). Such a transfer between distant grains
proceeds via a string of intermediate grains, where typically the
energy of a state with an extra electron is high ($\varepsilon\sim
E_C$). As in the case of a standard single-particle tunneling via
resonant impurity states, the entire process is realized as a
coherent sequence of local hops between adjacent grains in the
string. There is an important difference, however. For the
single-particle problem, where {\it the same} electron has to
tunnel sequentially through all the impurities in the string,
(starting from the first one and ending with the last), the order
of these local hops is fixed. In our case there are many electrons
in each grain, all these electrons being ready to tunnel to an
adjacent grain at any time, so that the sequence of local hops in
such a co-tunneling process can be {\it arbitrary}. As a result,
the states with a number of excited electrons and holes on
different grains appear as intermediate virtual states of the
cotunneling process, and the number of charged grains in these
intermediate states can be larger than one. In the final state,
however, all the charges of the intermediate grains should be
compensated, and the only extra charge is transferred between the
two terminal grains of the string.  Long-range hopping process
involves many intermediate grains; in general co-tunnelling
through some of them will be of elastic type, whereas some other
will be inelastic. We will see below that  elastic cotunneling
dominates the variable-range hopping at rather low temperatures
$T<T_c$, where $T_c$ is significantly lower than in the case of
single intermediate quantum dot.

Note also that in the case of not extremely low temperatures $T\gg
\delta$ (when the spectrum of electrons in grains can be treated
as a quasicontinuous one), one does not need to invoke phonons to
ensure the energy conservation: the energy can be taken from the
fermionic thermostat via inelastic cotunneling process.

{\it General approach to variable-range cotunneling.} The
Hamiltonian of the system has the form
$\hat{H}=\hat{H}_0+\hat{H}_{\rm tun}+\hat{H}_C$. Here
the"single-grain" hamiltonian $\hat{H}_0=\sum_{i}\hat{H}_0^{(i)}$,
and the inter-grain tunneling hamiltonian $\hat{H}_{\rm
tun}=\sum_{\langle ij\rangle}\hat{H}_{\rm tun}^{(ij)}$, in the
latter the summation runs over the pairs $\langle ij\rangle$ of
neighboring grains,
\begin{eqnarray*}
\hat{H}_0^{(i)}=\sum_{\alpha_i,\sigma}\epsilon_{\alpha_i}a^{+}_{\alpha_i\sigma}a_{\alpha_i\sigma},\quad
\hat{H}_{\rm
tun}^{(ij)}=\sum_{\alpha_i,\alpha_j,\sigma}t_{\alpha_i\alpha_j}a^{+}_{\alpha_i\sigma}a_{\alpha_j\sigma}
\end{eqnarray*}
where the operator $a^{+}_{\alpha_i\sigma}$ creates an electron in
a single-particle orbital eigenstate $\alpha_i$ with a spin
projection $\sigma$ on a grain $i$. The Coulomb interaction
hamiltonian
\begin{eqnarray}
\hat{H}_C=\frac12\sum_{ij}E_c^{(ij)}(\hat{n}_i-q_i)(\hat{n}_{j}-q_{j}),
\label{coulomb1}
\end{eqnarray}
$\hat{n}_i=\sum_{\alpha_i,\sigma}a^{+}_{\alpha_i\sigma}a_{\alpha_i\sigma}-n^{(0)}_i$
being  the  operator of excess number off electrons at the grain
$i$, $n^{(0)}_i$ corresponds to the minimum of
$\hat{H}_0+\hat{H}_C$. Variables $q_i$ (not necessarily integer!)
are the so called background charges (in the units of $e$). We
will treat them as independent continuous random variables
$-1/2<q_i<1/2$ with symmetric distribution $P(q)$. Two different
limits should be distinguished: (i) the case of strong charge
disorder, when the background charges are large, so that $q_i$ is
distributed homogeneously in the interval $-1/2<q_i<1/2$; and (ii)
the case of weak disorder, when the charges are small, so that
$q_0\equiv(\overline{q^2})^{1/2}\ll 1$ and the probability to have
$q=1/2$, related to the "bare density of ground states" at the
Fermi-level (cf. \cite{ZhangShklovskii}), is small: $P_{1/2}\equiv
P(1/2)\ll 1$. While the former case seems to be most appropriate
for a naturally disordered granular material, the latter one may
be relevant for high quality artificial arrays of quantum dots.

If the tunnel matrix elements $t_{\alpha_i\alpha_j}$ are small
enough, the rate $w_{ij}$ of  cotunneling between distant grains
$i$ and $j$ can be found in the high order perturbation theory in
$\hat{H}_{\rm tun}$:
\begin{widetext}
\begin{eqnarray}
w_{ij}=2\pi f_F(\varepsilon_i)f_F(-\varepsilon_j)\sum_{\cal
P}\sum_{\sigma_0,\{\sigma_l\}}
\sum_{\{p_l,h_l\}}\sum_{\{\alpha_m\}}\prod_{k=1}^{N+1}|t_{h_kp_{k-1}}|^2f_F(\epsilon_{p_0})f_F(-\epsilon_{h_{N+1}})
\prod_{l}f_F(\epsilon_{p_l})f_F(-\epsilon_{h_{l}})\nonumber\\
\times \delta\left[\epsilon_{p_0}-\epsilon_{h_{N+1}}+
\sum_{l}(\epsilon_{p_l}-\epsilon_{h_l})-\Delta_{ij}\right]\sum_{\cal
T,T'}(-1)^{K({\cal T})-K({\cal
T'})}\prod_{m}\delta_{\lambda_m({\cal T})\lambda_m({\cal
T'})}f_F\left[\lambda_m({\cal T})\epsilon_{\alpha_m}\right]
Q({\cal T})Q({\cal T'}) \label{amp1}
\end{eqnarray}
\end{widetext}

Here we have assumed that the $i\to j$ transition is dominated by
tunneling along a unique "string" -- a chain of neighboring
grains, denoted by numbers $k=0,1,\ldots ,N,N+1$, so that $0\equiv
i$ is the initial grain; $N+1\equiv j$ is the final grain; each
pair $k,k+1$ are in contact. The possibility for several relevant
strings to exist, and the effect of interference of their
contributions will be discussed in the last section of this
Letter. The summation runs over all possible partitions ${\cal P}$
of the string into two subsets: $\{k\}=\{m\}\cup\{l\}$; on the
grains $\{m\}$ the elastic cotunneling (via a state
$h_m,p_m\equiv\alpha_m$) occurs, while at grains $\{l\}$ the
inelastic cotunneling with a creation of an electron-hole pair
with quantum numbers $h_l,p_l$ takes place. The summation over all
eigenstates $\alpha_m$ and pairs  of eigenstates $h_l\neq p_l$ is
assumed. The energies $\epsilon_{pl},\epsilon_{hl}$ are measured
with respect to the Fermi level.  The spin-variable $\sigma_0$
corresponds to the state $p_0$; variables $\sigma_l$ correspond to
$p_l$-components of electron-hole pairs. All other spin-variables
are not independent because of the spin-conservation by tunneling
hamiltonian.  The spin summation gives the factor $2^{L({\cal
P})}$, where $L({\cal P})$ is a number of inelastic grains in the
partition ${\cal P}$.  The interference cross-terms between the
processes with different $\alpha_m$ are neglected because of
violent sign-fluctuations of $t_{h_kp_{k-1}}$.

The "time-orderings" ${\cal T}$ are all the possible orderings
$\{k_1,k_2,\ldots ,k_r, ,k_{N+1}\}$ of the set of grains
$k=0,\ldots ,N$ (note, that there is no final grain $N+1$ in this
set!). The contribution to the composite transition amplitude
$\hat{H}_{\rm eff}^{(ij)}$ corresponding to particular ${\cal T}$
has the structure
\begin{eqnarray*}
\hat{H}_{\rm eff}^{(ij)}({\cal T})=\hat{H}_{\rm
tun}^{(k_{N+1}k_{N+1}+1)}\hat{\cal G}\hat{H}_{\rm
tun}^{(k_{N}k_{N}+1)}\hat{\cal G}\cdots \hat{\cal G}\hat{H}_{\rm
tun}^{(k_{1}k_{1}+1)}
\end{eqnarray*}
where the many-particle Green-function $\hat{\cal
G}=[\hat{H}_0+\hat{H}_C]^{-1}$. The
 function $\lambda_k({\cal T})=1$, if in the ordering ${\cal T}$ the grain $k-1$
 comes
earlier, than the grain $k$, and $\lambda_k({\cal T})=-1$
otherwise. The sign-factor $(-1)^{K({\cal T})}$ arises due to
permutations of fermionic operators. The Fermi-functions
$f_F(\epsilon)$ take into account the Fermi filling factors. The
$\delta$-function ensures the energy conservation, and
$\Delta_{ij}=\varepsilon_j-\varepsilon_i-E_c^{(ij)}$ is the
difference of energies of the initial an final state. The factors
\begin{eqnarray}
Q({\cal T})= \prod_{r=1}^{N}\left[H_C\{n(r)\}+\sum_{r'=1}^{r}
(\epsilon_{h_{k_{r'}+1}}- \epsilon_{p_{k_{r'}}})\right]^{-1}
\label{amp1x}
\end{eqnarray}
are products of energy denominators, appearing in $\hat{\cal G}$.
Here $n_k$ are numbers of excess electrons on $k$-th grain after
$p$ local hops; they can be found from the following recursion
formula:
\begin{eqnarray}
n_k(r)=\left\{\begin{aligned} n_k(r-1)-1, & \qquad\mbox{if $\; k=k_r$},\\
n_k(r-1)+1, & \qquad\mbox{if $\; k=k_r+1$},\\
n_k(r-1), & \qquad\mbox{otherwise},
\end{aligned}\right.
\label{coulomb2}
\end{eqnarray}
while $n_k(0)=n_k^{(0)}$ is the equilibrium distribution.

Inspecting the expression \eqref{amp1} we see that the
characteristic value $\overline{\epsilon}_{\rm inel}$ of the
"inelastic energies" $\epsilon_{pl},\epsilon_{hl}$ is controlled
by the combination of the $\delta$-function in \eqref{amp1} and
the product of the corresponding Fermi-functions. As a result,
$\overline{\epsilon}_{\rm inel}\sim \Delta_{ij}/\overline{L}$,
where $\overline{L}$ (which is $T$-dependent) is the number of
inelastic cotunneling events in the total process. Actually $T\ll
\overline{\epsilon}_{\rm inel}\ll E_C$, so that, in particular,
the dependence of $Q({\cal T})$ on the inelastic energies can be
neglected. On the other hand, the characteristic value of
$\overline{\epsilon}_{\rm el}$ of the "elastic energies"
$\epsilon_{\alpha m}$  is limited only by the energy denominators
$Q({\cal T})$, so  that $\overline{\epsilon}_{\rm el}\sim E_C$.

Thus, performing the integration over
$\epsilon_{pl},\epsilon_{hl}$ and $\epsilon_{\alpha m}$, we obtain
for the effective Miller-Abrahams dimensionless conductance
$g_{ij}$ between two distant grains $i$ and $j$:
\begin{eqnarray}
g_{ij}\propto
e^{-\frac{\varepsilon_{ij}}{T}}\left(\frac{t}{E_C}\right)^{2N}\sum_{L=0}^N
\frac{\left(\frac{2|\Delta_{ij}|^2}{\delta^2}\right)^{L}\left(\frac{E_C}{\delta}\right)^{N-L}}{(2L+1)!}F_{NL}
\label{amp1q}
\end{eqnarray}
Here   $\delta=E_0e^{-\overline{\ln (E_0v\nu)}}$ is the mean level
spacing ($\nu$ is the electronic density of states {\it per one
spin projection} at the Fermi level in a particular grain, $v$ is
the grains volume, and $E_0$ is an arbitrary energy unit). The
characteristic Coulomb energy $E_C=E_0e^{\overline{\ln
(E_{kk}/E_0)}}$ (normally $E_C\sim e^2/\kappa_{\rm eff}a$, where
$a$ is the average diameter of grains, and $\kappa_{\rm eff}$ is
the effective dielectric permeability of the material, see
\cite{ZhangShklovskii}). Finally, he mean tunneling amplitude
$t=E_0e^{1/2\overline{\ln(|t_{kk+1}|^2/E_0^2)}}$, where
$|t_{kk+1}|^2$ is the "coarse grained" (i.e., averaged over an
interval of energies near the Fermi level, large compared to the
level spacing, but small, compared to any other relevant scale)
value of $|t_{\alpha_k,\alpha'_{k+1}}|^2$. The "averaging of the
logarithm" rule appearing in the above definitions of mean values,
arises as a result of self-averaging
 of large ($\sim N$) number of similar independent random factors
with identical distributions. Note also the presence of the
spin-factors 2 in the multipliers, corresponding to inelastic
processes and absence of such factors for the elastic processes.

 The local activation
energy $\varepsilon_{ij}$  for the $ij$ hop is the combination of
$\varepsilon_{i},\varepsilon_{j}$ and $E_c^{(ij)}$, standard for
the hopping conductivity theory  (see \cite{ES} for the explicit
definition).

The weight-function $F_{NL}=\sum_{\cal P} \delta_{L,L({\cal
P})}C({\cal P})$, where $C({\cal P})$ are numerical coefficients,
depending only on the partition ${\cal P}$ and on the explicit
form of the charging energy matrix $E_c^{(kk')}$:
\begin{eqnarray}
C({\cal P})=\sum_{\cal T,T'}(-1)^{K({\cal T})-K({\cal
T'})}\nonumber\\\times\int\prod_m
d\tilde{\epsilon}_m\delta_{\lambda_m({\cal T})\lambda_m({\cal
T'})}\theta\left[-\lambda_m({\cal
T})\tilde{\epsilon}_{\alpha_m}\right] \tilde{Q}({\cal
T})\tilde{Q}({\cal T'}) \label{amp1p}
\end{eqnarray}
$\tilde{\epsilon}_m\equiv\epsilon_m/E_C$ and $\tilde{Q}({\cal
T})\equiv Q({\cal T})E_C^{N}$ being the dimensionless variables.

The explicit form of the weight-function $F_{N,L}$ for general
$E_c^{(kk')}$ can not be found. However, as it is argued below,
the asymptotics of $F$, relevant for the purely elastic and purely
inelastic limits, are $F_{N,0}\approx\tilde{A}_1^N$ and
$F_{N,N}\approx\tilde{A}_2^N$, correspondingly. The numerical
constants $\tilde{A}_1$ and $\tilde{A}_2$ are not known. As a
result,
\begin{eqnarray}
g_{ij}\propto
\exp\left\{-\frac{\varepsilon_{ij}}{T}\right\}\left\{\begin{aligned}
\left(\frac{\tilde{A}_1g\delta}{8\pi^2E_C}\right)^{N_{ij}},&\quad\mbox{elastic,}\\
\left(\frac{e^2\tilde{A}_2g|\Delta_{ij}|^2}{16\pi^2N_{ij}^2E_C^2}\right)^{N_{ij}},&\quad\mbox{inelastic,}
\end{aligned}\right.
\label{resu2}
\end{eqnarray}
where $g\equiv Gh/e^2=8\pi^2(t/\delta)^2\ll 1$ is the average
dimensionless conductance of a contact between two adjacent
grains. Note, that this definition of $g$ differs from that in
Refs.\cite{ET02,FIS04,ANL}. Applying standard
Mott-Efros-Shklovskii arguments to the random network with
conductances \eqref{resu2}, we obtain Eq. \eqref{esl} with
$T_{ES}={\cal L}(T)E_C$
\begin{eqnarray}
 {\cal L}(T)=\left\{\begin{aligned}
c_1\ln\left(\frac{8\pi^2E_C}{\tilde{A}_1g\delta}\right),&\quad T\ll T_c,\\
c_1\ln\left(\frac{16\pi^2E_C^2}{e^2\tilde{A}_2gT^2{\cal
L}^2}\right),&\quad T\gg T_c,
\end{aligned}\right.
\label{resu3}
\end{eqnarray}
where the crossover temperature $T_c\sim\sqrt{E_C\delta}/{\cal
L}$, and  $c_1\sim 1$ is an unknown constant, depending on the
statistical  geometry of the granular material. Since $E_C\propto
a^{-1}$ and $\delta\propto a^{-3}$, we conclude that roughly
$T_{ES}\propto a^{-1}$ and $T_c\propto a^{-2}$.

It should be noted that the above consideration is justified and
the VRH regime is actual only if the characteristic length of the
hop is large: $\overline{N}\sim (E_C/{\cal L}T)^{1/2}\gg 1$. For
$\overline{N}< 1$ the Nearest Neighbor Hopping regime,
characterized by the Arrhenius law $\sigma\propto\exp(-E_A/T)$
with the activation energy $E_A\sim E_C$ should be observed. The
crossover temperature between the two regimes is controlled by the
intergrain conductance $g$; the NNH is likely to be found in the
samples with very low $g$.

 {\it Model of local
repulsion: Mott law for granular array.}
 The expression \eqref{amp1p} can be explicitly evaluated for the model
case of the short range Coulomb interaction
$E_c^{(kk')}=E_k\delta_{kk'}$. Then  dimensionless local energies
of the charged states are
$\tilde{E}_k^{(\pm)}=(E_k/E_C)\left[1/2\mp q_k\right]$, and
$\varepsilon_i=E_k\left(\frac12-|q_i|\right){\rm sign}(q_i)$. In
this case \eqref{amp1p} can be written as a product of single
particle Green-functions with energies, depending on the local
charge. As a result
\begin{eqnarray*}
C({\cal P})=\prod_m \left(\frac{1}{\tilde{E}_m^{(+)}}+
\frac{1}{\tilde{E}_m^{(-)}}\right)
\prod_l\left(\frac{1}{\tilde{E}_l^{(+)}}-
\frac{1}{\tilde{E}_l^{(-)}}\right)^2,
\end{eqnarray*}
so that, for $N\gg 1$, when the number of similar factors in
\eqref{amp1pa} is large and an effective self-averaging takes
place,
\begin{eqnarray}
F_{NL}=C_N^LA_1^{N-L}A_2^{L}\label{amp1pa}
\end{eqnarray}
\begin{eqnarray}
A_1\equiv e^{-\overline{\ln\left(1/4-q^2\right)}},\quad A_2\equiv
4e^{\overline{\ln
q^2}-2\overline{\ln\left(1/4-q^2\right)}},\label{average3}
\end{eqnarray}
$C_N^L$ being binomial coefficients. The constant $A_1$ does not
show any dramatic dependence on the strength of the random
potentials: $A_1=e^2\approx 7.4$ for  strong charge disorder, and
$A_1=4$ for a weak one. The constant $A_2=e^2$ for a strong
disorder, while for a weak one $A_1\sim q_0^2\ll 1$. The reason is
the destructive interference between two possible processes of the
pair production: in the "$e-h$-process" the electron is created
first and the hole is created the second, while in the
"$h-e$-process" the order is inverted. As a result, for $q_0\ll 1$
the crossover temperature strongly depends on $q_0$: $T_c(q_0)\sim
(q_0{\cal L})^{-1}\sqrt{E_C\delta}$. The growth of $T_c$ is
saturated at $T_c^{\rm max}\sim ({\cal
L})^{-1}(E_C^3\delta)^{1/4}$ for $q_0\lesssim(\delta /E_C)^{1/4}$,
when the energies $\epsilon_{pl},\epsilon_{hl}$ of the pairs come
into play. Thus, we conclude that for the case of weak charge
disorder the inelastic cotunneling is suppressed and the crossover
between elastic and inelastic cotunneling is shifted to higher
temperatures.

Unfortunately, the result \eqref{amp1pa} can not be generalized
for the case of nonlocal interaction $E_c^{(ij)}$. It can be
shown, however, that $F_{N,0}\approx \tilde{A}_1^N$, and
$F_{N,N}\approx \tilde{A}_2^N$, with certain renormalized
constants $\tilde{A}_1$ and $\tilde{A}_2$. Roughly, the reason is
as follows (details will be presented elsewhere):  simple
exponential form of the $F$-function holds for any "effectively
short-range" interaction (not necessarily strictly local one),
while for effectively long-range one the functional form of $F$
can be changed dramatically. The clue is that, despite the
long-range character of the Coulomb potential, the interaction of
effective degrees of freedom in our case is the short-range one.
Indeed, actual charge configurations, relevant to our problem, are
those, generated by local electronic hops between neighboring
grains. These hops create {\it local  dipoles}, and the
dipole-dipole interaction decays with distance $r$ as $r^{-3}$.

With the explicit formula \eqref{amp1pa} at hand one can perform
the summation in (\ref{amp1q}) and find
\begin{eqnarray*}
\ln g_{ij}= N_{ij}\left\{\ln\left(\frac{A_1 g\delta}{
8\pi^2E_{C}}\right)+\varphi\left(\frac{|\varepsilon_i-\varepsilon_j|}{2N_{ij}}
\sqrt{\frac{2A_2}{A_1E_{C}\delta}}\right)\right\}-
\frac{\varepsilon_{ij}}{T} ,
\end{eqnarray*}
where the function $\varphi(z)$ is implicitly defined by  the
relations $\varphi(z)=2y-\ln(1-y)$, $y^3=(1-y)z^2$. The function
$y(z)\equiv \overline{L}/N$ (with asymptotics $y(z)\approx
z^{2/3}$ at $z\ll 1$) has the meaning of a relative fraction of
inelastic events.

 The average number of grains $N_{ij}$ in a string,
connecting two distant grains, is proportional to the distance
$r_{ij}$ between them: $N_{ij}=c_2n_g^{1/d} r_{ij}$, where $n_g$
is the concentration of grains, and $c_2\sim 1$ is a geometric
constant, depending only on the statistics of grains packing.
Estimates, made for $c_2$ for several two-dimensional models show
that $c_2\approx 1$. Thus, we have arrived at the
$d+1$-dimensional percolation problem in the ${\bf r}_i$ and
$\varepsilon_i$ space. The density of sites
$\nu_{d+1}=n_gP_{1/2}/E_{C}$ in this space is the density of
marginal grains, whose ground states are almost degenerate. The
connectivity criterion reads
\begin{eqnarray}
\xi({\bf r}_i,\varepsilon_i|{\bf r}_j,\varepsilon_j)<\xi,\quad
\mbox{where}\quad \xi({\bf r}_i,\varepsilon_i|{\bf
r}_j,\varepsilon_j)=\frac{\varepsilon_{ij}}{T}+\nonumber\\
c_2n_g^{1/d} r_{ij}\left\{\ln\left(\frac{8\pi^2 E_{C}}{A_1
g\delta}\right)-\varphi\left(\frac{|\varepsilon_i-\varepsilon_j|}{2c_2n_g^{1/d}
r_{ij}}\sqrt{\frac{2A_2}{A_1E_{C}\delta}}\right)\right\}
\label{result7}
\end{eqnarray}
As usual (see \cite{ES}), one should find  a value $\xi=\xi_c$
corresponding to the first appearance of an infinite cluster of
grains, connected according to the criterion \eqref{result7}.
Then, with the exponential accuracy, the global conductivity of
the system $\sigma\propto\exp (-\xi_c)$.

The arising percolational model differs, however, from the
standard  VRH one (see \cite{ES,AHL}) due to an additional
dependence on $r_{ij}$ and $\varepsilon_i,\varepsilon_i$ appearing
in the argument of the function $\varphi$. However, since the
variation of this function on the relevant scale of it's argument
is $\delta\varphi\sim 1\ll {\cal L}$, the corresponding relative
variation of $\xi({\bf r}_i,\varepsilon_i|{\bf
r}_j,\varepsilon_j)$ is small  and  can be treated by the standard
perturbational method (see \cite{ES}). As a result, we obtain the
Mott law
\begin{eqnarray}
\sigma\sim \exp\{-(T_M/T)^{1/(d+1)}\}, \label{mott}
\end{eqnarray}
with $T_M$, which is  slightly temperature-dependent:
\begin{eqnarray} T_M=\frac{{\cal
L}^d(T/T_c)n_c}{2P_{1/2}}E_{C}, \quad T_c=\frac{2c_2}{\cal
L}\sqrt{\frac{A_1E_{C}\delta}{2A_2}},\label{result21}
\end{eqnarray}
\begin{eqnarray}
{\cal L}(T/T_c)=c_2\left\{\ln\left(\frac{8\pi^2 E_{C}}{A_1
g\delta}\right)-\chi(T/T_c)\right\} \label{result21e}
\end{eqnarray}
The universal percolation constant $n_c\approx 5.7$ for $d=3$ and
$n_c\approx 3.5$ for $d=2$. The function $\chi(z)$ is related to
$\varphi(z)$ by
\begin{eqnarray}
\chi(z)= \left\langle s\right\rangle^{-1}_{\rm perc}\left\langle
s\varphi\left(\frac{|\zeta-\zeta'|}{s}z \right)\right\rangle_{\rm
perc} ,\label{result18}
\end{eqnarray}
where the averaging  over the "percolation  hypersurface" in the
space of dimensionless energy $\zeta$ and dimensionless coordinate
${\bf s}$ has the following explicit meaning:
\begin{eqnarray*}
\left\langle F\right\rangle_{\rm perc}\equiv\frac {\int d\zeta
d\zeta' \int d{\bf s}F(\zeta,\zeta',s)
\delta\left(1-s-\frac{|\zeta|+|\zeta'|+|\zeta-\zeta'|}{2}\right)}
{\int d\zeta d\zeta'\int d{\bf s}
\delta\left(1-s-\frac{|\zeta|+|\zeta'|+|\zeta-\zeta'|}{2}\right)}
.
\end{eqnarray*}
The asymptotics of the function $\chi(z)$ are
\begin{eqnarray}
\chi(z)\approx\left\{\begin{aligned} b_1z^{2/3}\qquad &\mbox{for
$z\ll 1$},\\
2(\ln z+b_2+\ldots)\qquad & \mbox{for $z\gg 1$},
\end{aligned}
\right.  \label{result19}
\end{eqnarray}
where $b_1,b_2$ are universal constants, depending only on the
space dimensionality $d$. In particular, $b_2=-1/6$ for $d=2$ and
$b_2=-1/2$ for $d=3$.

Experimentally the Mott law \eqref{mott} is likely to be observed
in materials with weak charge disorder, where the density of
states at $\varepsilon=0$ is very small because of the factor
$P_{1/2}\ll 1$, and the Coulomb gap is irrelevant, except for
extremely low $T$ range. In such materials, however, $T_M$ is very
large (cf. \eqref{result21}) and the crossover between the Mott
and the Arrhenius laws should take place at relatively low $T$.

{\it Magnetoresistance.} In the entire above consideration we have
neglected a possible interference between contributions of
different  strings (if any), connecting the same pair $ij$. The
main reason for this approximation is the strong sign-fluctuations
of the matrix elements $t_{h_kp_{k-1}}$: even in the coherent
(purely elastic) limit the signs of contributions of different
strings  to the composite amplitude of $i\to j$ transition
fluctuate independently. The interference effects, although
irrelevant to the zero-field effects, are sensitive to magnetic
field, so that they may be the source for a low temperature effect
of a negative orbital magnetoresistance, similar to the one
discussed in \cite{magnet1,magnet2,magnet3,magnet4} for the case
of VRH in conventional disordered semiconductors. The key
component of this effect is the interference between the
contributions of different spatial paths, leading from $i$ to $j$.
In our case this means the existence of several strings $i\to j$
giving comparable contributions to $g_{ij}$. For a fairly
homogeneous material, where all $g_{kk+1}$ are of the same order
of magnitude one can expect such different strings to exist
already for $N_{ij}\gtrsim 1$.  For a strongly disordered material
(with exponentially large fluctuations of $g_{kk+1}$) there is
typically only one leading string for $N_{ij}<N_{\rm min}$, and
only for $N_{ij}>N_{\rm min}$ several strings act in parallel. The
crossover length $N_{\rm min}$ is a function of the magnitude of
fluctuations $D=\overline{(\ln g-\overline{\ln g})^2}$, the
explicit form of this function is model-dependent, and we will not
discuss it in the present Letter. For us it is only important that
$N_{\rm min}(D)\sim 1$ for $D\sim 1$, and $N_{\rm min}(D)\gg 1$
for $D\gg 1$.

Apparently, the magnetoresistance is controlled by the typical
area $S(N_{\rm loop})$ of the interference loop ($N_{\rm loop}$
being the "length" of this loop). Loops with $N_{\rm loop}<N_{\rm
min}$ are extremely rare, and can be ignored. For $N_{\rm
loop}>N_{\rm min}$ the scaling law $S(N_{\rm loop})\sim N_{\rm
loop}^{u}$, with an unknown exponent $u<2$ holds. The problem of
statistics of loops is closely related to the well-studied theory
of directed polymers in a random field \cite{polymers}.

\begin{figure}
\includegraphics[width=0.9\columnwidth]{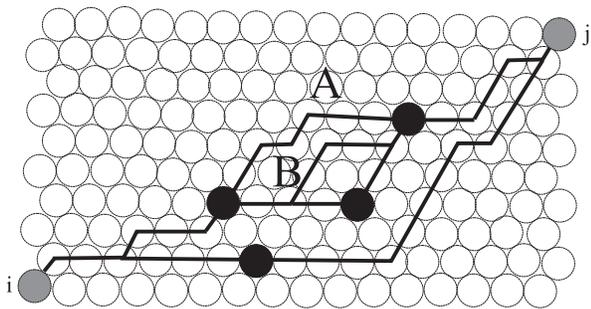}
\caption{Four different strings  contributing to the $i\to j$
transition in a particular realization of array. For a shown
partition ("inelastic grains" $\{l\}$ are depicted as filled
circles, elastic grains $\{m\}$ -- as open ones) only two strings
($A$ and $B$) contribute to interference effects.} \label{strings}
\end{figure}

 One of the essential ingredients
of our cotunneling process is the presence of inelastic events,
which certainly destroy the interference and suppress the
magnetoresistance. Namely, the interference between two different
strings $A$ and $B$ is possible only  if the cotunneling at all
grains of $A$ and $B$, which are not common for them, is elastic
(see Fig.\ref{strings}). Since the relative fraction of inelastic
cotunnelings $y$ depends on the temperature, so does the length
$M_{\rm el}\sim 1/y$ of a typical stretch on a string, containing
only "elastic grains". It is just $M_{\rm el}$, not the entire
length $N$ of the distant hop, that should play the role of the
effective length $N_{\rm loop}$ of the interference loop. Clearly,
for $T\gtrsim T_c$ one has $M_{\rm el}\sim 1$, while $M_{\rm
el}\sim (T_c/T)^{2/3}\gg 1$ for $T\ll T_c$.

Thus, we can conclude that for temperatures $T>T_{\rm mag}\sim
T_cN_{\rm min}^{-3/2}(D)<T_c$ the orbital magnetoresistance is
strongly suppressed (since the typical elastic stretch is shorter
than $N_{\rm min}$), while at $T\ll T_{\rm mag}$ it can be
relatively strong: the characteristic magnetic field $H_c$, at
which the conductivity would saturate at $\sigma (H\gg H_c)\sim
\sqrt{2}\sigma (H=0)$ (cf. Ref.\cite{magnet3}), is
\begin{eqnarray}
H_c\sim \Phi_0/S(M_{\rm el})\sim
H_c^{(0)}(T/T_c)^{2u/3},\label{scaling1}
\end{eqnarray}
where $H_c^{(0)}=\Phi_0n_g^{2/d}$ is the field, corresponding to a
flux $\Phi_0$ through an elementary triangle of neighboring
grains. The dependence $\sigma (H)$ at $H<H_c$ can be different in
different ranges of $H$: either $\Delta\sigma\propto H^2$ (at the
smallest fields, see \cite{magnet2,magnet4}) or
$\Delta\sigma\propto H$ (at the intermediate fields, see
\cite{magnet1,magnet4}), or $\Delta\sigma\propto H^{1/2}$ (at
relatively high fields, see \cite{magnet3}).

{\it Conclusions.} In conclusion, we have developed a theory of
variable-range hopping in granular arrays with poor intergrain
coupling. Long-range hopping of electrons is provided by the
multiple co-tunnelling "strings" which contains both elastic and
inelastic processes within individual grains.  In the presence of
long-range Coulomb interaction, Efros-Shklovskii law for the
temperature dependence of conductivity is derived in the
asymptotic limits of purely inelastic or purely elastic
co-tunnelling.
 Upon temperature decrease, relative contribution of elastic co-tunnelling
increases; For the model case of local (screened) Coulomb
interaction, general situation of partially elastic co-tunnelling
was studied and crossover temperature $T_c$ was determined, cf.
Eqs.(\ref{resu3},\ref{result21}). For real granular metals this
crossover temperature happens to be rather low. In particular, for
Al grains of size $a\sim 20 nm$  one estimates $E_C \sim 500 K$
and $\delta \sim 0.05 K$, which leads (at $g\sim 0.3$, so that
${\cal L}\sim 12$) to $T_c \sim 0.5 K$. For $a\sim 10 nm$ the same
estimates give $T_c \sim 2 K$. Therefore the major part of
experimental temperature range (from room to liquid helium
temperatures) is dominated by inelastic co-tunnelling. This is the
reason for magnetoresistance to be very weak in granular metals,
contrary to disordered semiconductors. Observation of noticeable
negative magnetoresistance due to interference of different
tunnelling "strings" might be possible with granular media made of
small ($\leq 10 $ nm) grains of non-superconductive metals like
copper, silver or gold,  at  temperatures below $1 K$. In the
inelastic regime (at $T>T_c$) the "constant" $T_{ES}$, entering
Eq\eqref{esl}, is itself $T$-dependent: it logarithmically
increases with the decrease of $T$. This dependence should lead to
somewhat faster growth of the resistivity upon lowering $T$, than
predicted by the standard Efros-Shklovskii law.

After the present study was completed, we became aware of the
preprint~\cite{ANL} where purely elastic variable-range
cotunnelling was proposed as the conduction mechanism for granular
metals; their results seem to agree with our ones as long as
inelastic processes are neglected.

We are grateful to A.L.Efros, T.Grenet, L.B.Ioffe and Z.Ovadyahu
for many useful discussions. This research was supported by
Programm "Quantum Macrophysics" of Russian Academy of Sciences and
by RFBR grants \# 04-02-16348 and \# 04-02-08159.


\begin{thebibliography}{99}

\bibitem{exp1} T.Chui {\it et al}, Phys. Rev. B {\bf 23}, 6172 (1981).

\bibitem{exp2} R.W.Simon {\it et al}, Phys. Rev. B {\bf 36}, 1962 (1987).

\bibitem{exp3} A.Gerber, {\it et al}, Phys. Rev. Lett., {\bf 78},
4277 (1997).

\bibitem{EfrosShklovskii75}  A.L.Efros and B.I.Shklovskii, J.Phys. {\bf C 8}, L49 (1975).

\bibitem{ES} B.I.Shklovskii and
A.L.Efros, Electronic properties of doped semiconductors,
Springer-Verlag, NY, 1984.

\bibitem{ZhangShklovskii} J.Zhang and B.I.Shklovskii, Phys.Rev.{\bf
B 70}, 115317 (2004)

\bibitem{AverinNazarov90}  D.A.Averin and Yu.V.Nazarov, Phys. Rev. Lett., {\bf 65}, 2446
(1990).

\bibitem{ET02} K.B.Efetov and A.Tschersich, Europhys. Lett., {\bf
59}, 114 (2002); Phys. Rev. {\bf B 67}, 174205 (2003).

\bibitem{FIS04}  M.V.Feigelman, A.S.Ioselevich and M.A.Skvortsov,
Phys. Rev. Lett. {\bf 93}, 136403 (2004).



\bibitem{magnet1} V.L.Nguen, B.Z.Spivak and B.I.Shklovskii, JETP Lett., {\bf 41}, 42
(1985).

\bibitem{magnet2}
U.Sivan, O.Entin-Wohlman and Y.Imry, Phys.Rev.Lett., {\bf 60},
1566 (1988)

\bibitem{magnet3}
H.L.Zhao, B.Z.Spivak, M.P.Celfand and S.Feng, Phys.Rev. {\bf B
44}, 10760 (1991)

\bibitem{magnet4}
B.I.Shklovskii and  B.Z.Spivak, In: Hopping transport in solids,
ed. by M.Pollak and B.Shklovskii, (Elsevier, Amsterdam, 1991,
p.271.



\bibitem{GlazmanReview}  L.I.Glazman and M.Pustilnik, cond-mat/0501007

\bibitem{orthodox} D.V.Averin and K.K.Likharev,
In: {\it Mesoscopic Phenomena in Solids}, ed. by B.L.Altshuler,
P.A.Lee, and R.A.Webb (Elsevier, Amsterdam, 1991).

\bibitem{AHL} V.Ambegokar, B.I.Halperin, and J.S.Langer, Phys.Rev.{\bf
B 4}, 2612 (1971).

\bibitem{polymers} See, e.g., T.Hwa and D.S.Fisher, Phys.Rev. {\bf B 49}, 3136
(1994), and references therein.

\bibitem{ANL}  I. S. Beloborodov, A.V.Lopatin, V.M.Vinokur, and V.I.Kozub,
cond-mat/0501094.


\end{thebibliography}
\end{document}